\documentclass[preprint,double-spaced,showpacs,preprintnumbers,amsmath,amssymb]{revtex4}

\usepackage{graphicx}
\usepackage{dcolumn}
\usepackage{bm,slashed}
\usepackage{amssymb}

\begin{document}

\preprint{}

\title{Towards differential elimination of spinor field from spinor electrodynamics}

\author{Andrey Akhmeteli}
 \email{akhmeteli@ltasolid.com}
 \homepage{http://www.akhmeteli.org}
\affiliation{%
LTASolid Inc.\\
10616 Meadowglen Ln 2708\\
Houston, TX 77042, USA}


\date{\today}

\begin{abstract}
A system of PDEs for the electromagnetic field and one real component of the spinor field is generally equivalent to spinor electrodynamics. There are reasons to believe that the component can be differentially eliminated from the system.

A Lagrangian depending on the electromagnetic field and one real component of the spinor field generally describes the same physics as spinor electrodynamics.
\end{abstract}

\maketitle

\section{\label{sec:level1}Introduction}

Spinor electrodynamics (Dirac-Maxwell electrodynamics) is a reasonably realistic and important theory as its quantization yields quantum electrodynamics.

Some recent results suggest a possibility of differential elimination of the spinor field from the equations of spinor electrodynamics. On the one hand, the matter field can be algebraically eliminated from scalar electrodynamics (Klein-Gordon-Maxwell electrodynamics) in the unitary gauge, the resulting equations describe independent evolution of electromagnetic field, and can be naturally embedded into a quantum field theory ~\cite{Akhm10,Akhmeteli-IJQI,Akhmeteli-EPJC}. On the other hand, similar results can be obtained for spinor electrodynamics, but at the cost of introduction of a complex 4-potential of electromagnetic field ~\cite{Akhmeteli-EPJC}, which is less satisfactory.

In the following, the results of ~\cite{Akhmeteli-JMP} are used, where it was shown that, in a general case, three out of four complex components of the Dirac spinor function can be algebraically eliminated from the Dirac equation, and the remaining component can be made real (at least locally) by a gauge transform. In Section II, the equations of spinor electrodynamics are rewritten as five real partial differential equations (PDE) for the 4-potential of electromagnetic field and the only real component of the spinor field $\psi_1$. The equations contain derivatives of $\psi_1$ of up to the third order. According to simple and probably naive calculations, taking all derivatives of up to the fourth order of these equations with respect to the four spacetime coordinates may yield an algebraic system of polynomial equations (with respect to derivatives of $\psi_1$ of up to the seventh order) that contains more equations than indeterminates. Therefore, hopefully, $\psi_1$ can be expressed via the 4-potential of the electromagnetic field and its derivatives using this system of equations and eliminated from the equations of spinor electrodynamics. The system is very large, so it is not clear if this elimination can be actually performed even using state-of-the-art hardware and software. It is also difficult to say if the resulting equations will describe independent evolution of electromagnetic field, but one can hope so based on the results for scalar electrodynamics.

So what could be the benefits of elimination of the spinor field from spinor electrodynamics? It would reveal that modified Maxwell equations can describe both the electromagnetic and spinor field (cf. ~\cite{Dirac}), allow natural embedding of spinor electrodynamics in a quantum field theory (cf. ~\cite{Akhmeteli-EPJC}) and possibly suggest a new approach to quantization of spinor electrodynamics. One can also speculate that the structure of solutions of the system of polynomial equations may be quite different for some specific value of the fine-structure constant $\alpha$, which could provide a hint on the theoretical value of this constant.

In Section III, a Lagrangian of spinor electrodynamics is derived where the spinor field is described by just one real component. This is not trivial, as elimination of the three components of the spinor from the Dirac equation and making the remaining component real by a gauge transform actually yields two real equations for the real component. However, one of the equations is equivalent to the current conservation equation, which follows also from the Maxwell equation. The Lagrangian derivation is similar to that for scalar electrodynamics, where the scalar matter field can be made real by a gauge transform ~\cite{Schroed}, and a Lagrangian ~\cite{Takabayasi} can be derived where the charged matter field is described by one real function ~\cite{Akhmeteli-EPJC,Akhm18}. Furthermore, the matter field can be eliminated from the Lagrangian, and the resulting Lagrangian generally describes the same physics as scalar electrodynamics, but depends only on the electromagnetic field ~\cite{Akhm18}.

\maketitle

\section{\label{sec:level1}An approach to elimination of the spinor field from the equations of spinor electrodynamics}
The equations of spinor electrodynamics are as follows:
\begin{equation}\label{eq:pr25}
(i\slashed{\partial}-\slashed{A}-1)\psi=0,
\end{equation}
\begin{equation}\label{eq:pr25a}
\bar{\psi}(i\overleftarrow{\slashed{\partial}}+\slashed{A}+1)=0,
\end{equation}
\begin{equation}\label{eq:pr26}
\Box A_\mu-A^\nu_{,\nu\mu}=e^2\bar{\psi}\gamma_\mu\psi,
\end{equation}
where, e.g., $\slashed{A}=A_\mu\gamma^\mu$ (the Feynman slash notation), and a derivative with a left arrow above acts to the left. For the sake of simplicity, a system of units is used where $\hbar=c=m=1$, and the electric charge $e$ is included in $A_\mu$ ($eA_\mu \rightarrow A_\mu$).
The chiral representation of $\gamma$-matrices ~\cite{Itzykson}
\begin{equation}\label{eq:d1}
\gamma^0=\left( \begin{array}{cc}
0 & -I\\
-I & 0 \end{array} \right),\gamma^i=\left( \begin{array}{cc}
0 & \sigma^i \\
-\sigma^i & 0 \end{array} \right)
\end{equation}
is used, where index $i$ runs from 1 to 3, and $\sigma^i$ are the Pauli matrices.

The spinor field $\psi$ and the electromagnetic field $A^\mu$ depend on the spacetime coordinates $x=(x^0,x^1,x^2,x^3)$, so, for example, $\psi_{,\mu}=\frac{\partial \psi}{\partial x^\mu}$. The metric tensor used to raise and lower indices is ~\cite{Itzykson}
\begin{equation}\label{eq:d1ps}
\nonumber
g_{\mu\nu}=g^{\mu\nu}=\left( \begin{array}{cccc}
1 & 0 & 0 & 0\\
0 & -1 & 0 & 0\\
0 & 0 & -1 & 0\\
0 & 0 & 0 & -1 \end{array} \right), g_\mu^\nu=\delta_\mu^\nu.
\end{equation}
Now, to fix the gauge, let us require that $\psi_1$, the first component of the spinor field
\begin{equation}\label{eq:d2}
\psi=\left( \begin{array}{c}
\psi_1\\
\psi_2\\
\psi_3\\
\psi_4\end{array}\right),\bar{\psi}=\psi^\dagger\gamma_0= (-\psi_3^*,-\psi_4^*,-\psi_1^*,-\psi_2^*),
\end{equation} is real: $\psi_1=\psi_1^*$. As $\psi_1$ generally does not vanish identically, this does fix the gauge (in a general case).

Now let us eliminate all spinor components but $\psi_1$ from the equations of spinor electrodynamics (\ref{eq:pr25},\ref{eq:pr25a},\ref{eq:pr26}) under the gauge condition.

We will use the elimination process of  work ~\cite{Akhmeteli-JMP,AkhmeteliJMParx}. The Dirac equation (\ref{eq:pr25}) can be written in components as follows:
\begin{eqnarray}\label{eq:d3}
(A^0+A^3)\psi_3+(A^1-i A^2)\psi_4
+i(\psi_{3,3}-\imath\psi_{4,2}+\psi_{4,1}-\psi_{3,0})=\psi_1,
\end{eqnarray}
\begin{eqnarray}\label{eq:d4}
(A^1+i A^2)\psi_3+(A^0-A^3)\psi_4
-i(\psi_{4,3}-i\psi_{3,2}-\psi_{3,1}+\psi_{4,0})=\psi_2,
\end{eqnarray}
\begin{eqnarray}\label{eq:d5}
(A^0-A^3)\psi_1-(A^1-i A^2)\psi_2
-i(\psi_{1,3}-i\psi_{2,2}+\psi_{2,1}+\psi_{1,0})=\psi_3,
\end{eqnarray}
\begin{eqnarray}\label{eq:d6}
-(A^1+i A^2)\psi_1+(A^0+A^3)\psi_2
+i\psi_{2,3}+\psi_{1,2}-i(\psi_{1,1}+\psi_{2,0})=\psi_4.
\end{eqnarray}
Obviously, equations (\ref{eq:d5},\ref{eq:d6}) can be used to express components $\psi_3,\psi_4$ via $\psi_1,\psi_2$ and eliminate them from equations (\ref{eq:d3},\ref{eq:d4}) (cf. \cite{Feygel}). The resulting equations for $\psi_1$ and $\psi_2$ are as follows:
\begin{eqnarray}\label{eq:d7}
\nonumber
-\psi_{1,\mu}^{,\mu}+\psi_2(-i A^1_{,3}-A^2_{,3}+A^0_{,2}+A^3_{,2}+
i(A^0_{,1}+A^3_{,1}+A^1_{,0})+A^2_{,0})+\\
\psi_1(-1+A^{\mu} A_{\mu}-i A^{\mu}_{,\mu}+i A^0_{,3}-A^1_{,2}+A^2_{,1}+i A^3_{,0})-2 i A^{\mu}\psi_{1,\mu}=0,
\end{eqnarray}
\begin{eqnarray}\label{eq:d8}
\nonumber
-\psi_{2,\mu}^{,\mu}+i\psi_1( A^1_{,3}+i A^2_{,3}+i A^0_{,2}-i A^3_{,2}+
A^0_{,1}-A^3_{,1}+A^1_{,0}+i A^2_{,0})+\\
\psi_2(-1+A^{\mu} A_{\mu}-i( A^{\mu}_{,\mu}+A^0_{,3}+i A^1_{,2}-i A^2_{,1}+ A^3_{,0}))-2 i A^{\mu}\psi_{2,\mu}=0.
\end{eqnarray}

As equation (\ref{eq:d7}) contains $\psi_2$, but not its derivatives, it can be used to express $\psi_2$ via $\psi_1$:
\begin{eqnarray}\label{eq:d8nn1}
\psi_2=-\left(i F^1+F^2\right)^{-1}\left(\Box'+i F^3\right)\psi_1,
\end{eqnarray}
where $F^i=E^i+i H^i$, electric field $E^i$ and magnetic field $H^i$ are defined by the standard formulae
\begin{eqnarray}\label{eq:d8n1}
F^{\mu\nu}=A^{\nu,\mu}-A^{\mu,\nu}=\left( \begin{array}{cccc}
0 & -E^1 & -E^2 & -E^3\\
E^1 & 0 & -H^3 & H^2\\
E^2 & H^3 & 0 & -H^1\\
E^3 &-H^2 & H^1 & 0  \end{array} \right),
\end{eqnarray}
and the modified d'Alembertian $\Box'$ is defined as follows:
\begin{eqnarray}\label{eq:d8n2}
\Box'=\partial^\mu\partial_\mu+2i A^\mu\partial_\mu+i A^\mu_{,\mu}-A^\mu A_\mu+1.
\end{eqnarray}
Using the above notation, equation (\ref{eq:d8}) can be rewritten as follows:
\begin{eqnarray}\label{eq:d8nn2}
-\left(\Box'-i F^3\right)\psi_2-\left(i F^1-F^2\right)\psi_1=0,
\end{eqnarray}
so equation (\ref{eq:d8nn1}) can be used to eliminate $\psi_2$ from equation (\ref{eq:d8nn2}), yielding an equation of the fourth order for $\psi_1$:
\begin{eqnarray}\label{eq:d8new}
\left(\left(\Box'-i F^3\right)\left(i F^1+F^2\right)^{-1}\left(\Box'+i F^3\right)-i F^1+F^2\right)\psi_1=0.
\end{eqnarray}

Note that this complex equation contains two real equations for  the real component $\psi_1$. In particular, these two equations imply current conservation. The current conservation can be written in the following form ~\cite{Akhmeteli-JMP,AkhmeteliJMParx}:
\begin{equation}\label{eq:d13new}
\textrm{Im}(\psi_4^* \delta)=0,
\end{equation}
where $\delta$ is the left-hand side of equation (\ref{eq:d8new}).

Obviously, equation (\ref{eq:d8new}) also implies the following equation:
\begin{equation}\label{eq:d13a}
\textrm{Im}(\left(i F^1+F^2\right) \delta)=0.
\end{equation}
One can check that this equation is a PDE of the third order with respect to $\psi_1$.

In a general case, $\textrm{Im}(\frac{i F^1+F^2}{\psi_4^*})$ does not vanish identically, so the system containing equations (\ref{eq:d13new},\ref{eq:d13a}) is equivalent to equation (\ref{eq:d8new}). On the other hand, current conservation (\ref{eq:d13new}) follows also from the Maxwell equation (\ref{eq:pr26}). Thus, if we eliminate all spinor components except $\psi_1$ from equation (\ref{eq:pr26}) using equations (\ref{eq:d5},\ref{eq:d6},\ref{eq:d8nn1}) and add equation (\ref{eq:d13a}), we will obtain a system (A) of five real PDEs of the third order with respect to the real indeterminate $\psi_1$, and this system of equations is generally equivalent to the equations of spinor electrodynamics (\ref{eq:pr25},\ref{eq:pr25a},\ref{eq:pr26}) under the gauge condition.

It is well-known that, if there are $d$ independent variables, the number of all different derivatives of order $r$, where $0\leq r \leq m$, equals the number of combinations
\begin{equation}\label{eq:dcom}
\left( \begin{array}{c} m+d\\ d\end{array}\right)=\frac{(m+d)!}{m!d!}.
\end{equation}
In our case, $d=4$ (the number of dimensions of spacetime). If we apply all different derivatives of order $r$, where $0\leq r \leq 4$, to the equations of system (A), we will obtain a system (B) containing
\begin{equation}\label{eq:dcom2}
5\frac{(4+4)!}{4!4!}=350
\end{equation}
equations. As the equations of system (A) are of the third order with respect to $\psi_1$, system (B) will contain, in the worst case, all different derivatives of $\psi_1$ of order $r$, where $0\leq r \leq 7$. Therefore, system (B) will contain no more than
\begin{equation}\label{eq:dcom3}
\frac{(7+4)!}{7!4!}=330
\end{equation}
such derivatives, which we can regard as indeterminates of a system of polynomial equations. Therefore, there are more equations than indeterminates in system (B), so, hopefully, it is possible to use the system to express $\psi_1$ via components of the 4-potential of electromagnetic field and their derivatives and thus eliminate the spinor field from spinor electrodynamics. The above estimates may be naive, and more rigorous estimates, such as those of ~\cite{Gust} may be required.

\section{\label{sec:level1}Lagrangian of spinor electrodynamics with just one real function to describe charged spinor field}
The Lagrangian of spinor electrodynamics is ~\cite{Bogo,Itzykson}
\begin{equation}\label{eq:lag}
\mathcal{L}=\frac{i}{2}\left(\bar{\psi}\gamma^\mu\psi_{,\mu}-\bar{\psi}_{,\mu}\gamma^\mu\psi\right)-\bar{\psi}\psi-\frac{1}{4 e^2}F_{\mu\nu}F^{\mu\nu}-A_\mu\bar{\psi}\gamma^\mu\psi.
\end{equation}
Let us derive the equations of motion for Lagrangian
\begin{eqnarray}\label{eq:lagrprim}
\nonumber
\mathcal{L}'=\mathcal{L}-i\lambda\left(\psi_1^*-\psi_1\right)=\\
\frac{1}{2}\bar{\psi}(i\gamma^\mu\psi_{,\mu}-A_\mu\gamma^\mu\psi-\psi)-
\frac{1}{2}(i\bar{\psi}_{,\mu}\gamma^\mu+A_\mu\bar{\psi}\gamma^\mu+\bar{\psi})\psi-\frac{1}{4 e^2}F_{\mu\nu}F^{\mu\nu}-i\lambda(\psi_1^*-\psi_1),
\end{eqnarray}
where the Lagrange multiplier $\lambda=\lambda(x)$ is real, under the condition $\psi_1^*-\psi_1$=0. It is not difficult to obtain:
\begin{equation}\label{eq:pr25d}
(i\slashed{\partial}-\slashed{A}-1)\psi-i\gamma^0\Lambda=0,
\end{equation}
\begin{equation}\label{eq:pr25ad}
\bar{\psi}(i\overleftarrow{\slashed{\partial}}+\slashed{A}+1)-i\Lambda^T=0,
\end{equation}
where
\begin{equation}\label{eq:lambdad2}
\Lambda=\left( \begin{array}{c}
\lambda\\
0\\
0\\
0\end{array}\right),\Lambda^T= (\lambda,0,0,0),
\end{equation}
and the Maxwell equation (\ref{eq:pr26}) does not change.

Let us add equation (\ref{eq:pr25d}) multiplied by $\bar{\psi}$ from the left and equation (\ref{eq:pr25ad}) multiplied by $\psi$ from the left:
\begin{equation}\label{eq:sum}
i\bar{\psi}\overleftarrow{\slashed{\partial}}\psi+i\bar{\psi}\slashed{\partial}\psi-
i\lambda(\psi_1^*+\psi_1)=0.
\end{equation}
On the other hand,
\begin{equation}\label{eq:curpres}
i\bar{\psi}\overleftarrow{\slashed{\partial}}\psi+i\bar{\psi}\slashed{\partial}\psi=
(\bar{\psi}\gamma^\mu\psi)_{,\mu}=0,
\end{equation}
as the Maxwell equation (\ref{eq:pr26}) implies current preservation, and, according to the gauge condition, $\psi_1^*=\psi_1$, so $\lambda\psi_1=0$. As in a general case $\psi_1$ does not vanish identically, we obtain $\lambda=0$. Therefore, Lagrangian $\mathcal{L}'$ (\ref{eq:lagrprim}) under the condition $\psi_1^*-\psi_1$=0 implies the same Dirac equation and Maxwell equation as Lagrangian $\mathcal{L}$ (\ref{eq:lag}) of spinor electrodynamics, but in a fixed gauge. Therefore, for every solution of the equations of spinor electrodynamics there is a physically equivalent solution for Lagrangian $\mathcal{L}'$ with the gauge condition (a solution that only differs in the choice of gauge). It is also obvious that each solution for Lagrangian $\mathcal{L}'$  under the gauge condition is a solution for Lagrangian $\mathcal{L}$. Thus, Lagrangian $\mathcal{L}$ and Lagrangian $\mathcal{L}'$ under the gauge condition describe the same physics.

Now let us eliminate all spinor components but $\psi_1$ from Lagrangian $\mathcal{L}'$  under the gauge condition using the Dirac equation and $\lambda=0$. Let us note that the second term in \ref{eq:lagrprim} is a complex conjugation of the first term, so let us focus on the first term $\frac{1}{2}\bar{\psi}(i\gamma^\mu\psi_{,\mu}-A_\mu\gamma^\mu\psi-\psi)$.

Equation (\ref{eq:d8}) can be rewritten as follows:
\begin{eqnarray}\label{eq:d8nn2}
-\left(\Box'-i F^3\right)\psi_2-\left(i F^1-F^2\right)\psi_1=0,
\end{eqnarray}
so equation (\ref{eq:d8nn1}) can be used to eliminate $\psi_2$ from equation (\ref{eq:d8nn2}), yielding an equation of the fourth order for $\psi_1$:
\begin{eqnarray}\label{eq:d8n3}
\left(\left(\Box'-i F^3\right)\left(i F^1+F^2\right)^{-1}\left(\Box'+i F^3\right)-i F^1+F^2\right)\psi_1=0.
\end{eqnarray}
Using (\ref{eq:d5},\ref{eq:d6},\ref{eq:d8nn1},\ref{eq:d8nn2}), one can see that all components of $i\gamma^\mu\psi_{,\mu}-A_\mu\gamma^\mu\psi-\psi$ vanish except for the second one, which equals the left-hand side of (\ref{eq:d8n3}). As the second component of $\bar{\psi}$ is $-\psi_4^*$ and
\begin{eqnarray}\label{eq:d8q}
-\psi_4^*=(A^1-i A^2-\partial_2-i \partial_1)\psi_1^*-(A^0+A^3-i\partial_3+i\partial_0)\psi_2^*
\end{eqnarray}
(cf. (\ref{eq:d6})), we obtain the following Lagrangian of spinor electrodynamics with just one real function to describe charged spinor field:
\begin{eqnarray}\label{eq:d8q2}
\nonumber
\mathcal{L}''=
\\
\nonumber
\textrm{Re}(((A^1-i A^2-\partial_2-i \partial_1-(A^0+A^3-i\partial_3+i\partial_0)(i F^{1*}-F^{2*})^{-1}(\Box''-i F^{3*}))\psi_1)\times
\\
((\Box'-i F^3)(i F^1+F^2)^{-1}(\Box'+i F^3)-i F^1+F^2)\psi_1)-\frac{1}{4 e^2}F_{\mu\nu}F^{\mu\nu},
\end{eqnarray}
where $\psi_1$ is real and
\begin{eqnarray}\label{eq:d8n2c}
\Box''=\partial^\mu\partial_\mu-2i A^\mu\partial_\mu-i A^\mu_{,\mu}-A^\mu A_\mu+1.
\end{eqnarray}
Simplification of the Lagrangian and derivation of its relativistically covariant form along the lines of ~\cite{Akhm2015} is left for future work.
\maketitle

\section{\label{sec:level1}Conclusion}

A system of five PDEs for the 4-potential of the electromagnetic field and one real component of the spinor field is generally equivalent to spinor electrodynamics. The equations are of the third order with respect to the component of the spinor field, and there are reasons to believe that the component can be differentially eliminated from the system. It is not clear, however, if this elimination can be actually performed using state-of-the-art hardware and software or if the resulting equations will describe independent evolution of electromagnetic field.

A Lagrangian depending on the 4-potential of the electromagnetic field and one real component of the spinor field generally describes the same physics as spinor electrodynamics.

\section*{Acknowledgments}

The author is grateful to V. G. Bagrov, A. V. Gavrilin, A. Yu. Kamenshchik, and nightlight for their interest in this work and valuable remarks.


\begin{thebibliography}{14}
\expandafter\ifx\csname natexlab\endcsname\relax\def\natexlab#1{#1}\fi
\expandafter\ifx\csname bibnamefont\endcsname\relax
  \def\bibnamefont#1{#1}\fi
\expandafter\ifx\csname bibfnamefont\endcsname\relax
  \def\bibfnamefont#1{#1}\fi
\expandafter\ifx\csname citenamefont\endcsname\relax
  \def\citenamefont#1{#1}\fi
\expandafter\ifx\csname url\endcsname\relax
  \def\url#1{\texttt{#1}}\fi
\expandafter\ifx\csname urlprefix\endcsname\relax\def\urlprefix{URL }\fi
\providecommand{\bibinfo}[2]{#2}
\providecommand{\eprint}[2][]{\url{#2}}

\bibitem[{\citenamefont{Akhmeteli}({\natexlab{a}})}]{Akhm10}
\bibinfo{author}{\bibfnamefont{A.~M.} \bibnamefont{Akhmeteli}},
  \emph{\bibinfo{title}{quant-ph/0509044}}.

\bibitem[{\citenamefont{Akhmeteli}(2011{\natexlab{a}})}]{Akhmeteli-IJQI}
\bibinfo{author}{\bibfnamefont{A.}~\bibnamefont{Akhmeteli}},
  \bibinfo{journal}{Int. J. Quantum Inf.} \textbf{\bibinfo{volume}{9}},
  \bibinfo{pages}{Suppl. 17} (\bibinfo{year}{2011}{\natexlab{a}}).

\bibitem[{\citenamefont{Akhmeteli}(2013)}]{Akhmeteli-EPJC}
\bibinfo{author}{\bibfnamefont{A.}~\bibnamefont{Akhmeteli}},
  \bibinfo{journal}{Eur. Phys. J. C} \textbf{\bibinfo{volume}{73}},
  \bibinfo{pages}{2371} (\bibinfo{year}{2013}).

\bibitem[{\citenamefont{Akhmeteli}(2011{\natexlab{b}})}]{Akhmeteli-JMP}
\bibinfo{author}{\bibfnamefont{A.}~\bibnamefont{Akhmeteli}},
  \bibinfo{journal}{J. Math. Phys} \textbf{\bibinfo{volume}{52}},
  \bibinfo{pages}{082303} (\bibinfo{year}{2011}{\natexlab{b}}).

\bibitem[{\citenamefont{Dirac}(1951)}]{Dirac}
\bibinfo{author}{\bibfnamefont{P.~A.~M.} \bibnamefont{Dirac}},
  \bibinfo{journal}{Proc. Roy. Soc. London A} \textbf{\bibinfo{volume}{209}},
  \bibinfo{pages}{291} (\bibinfo{year}{1951}).

\bibitem[{\citenamefont{Schr\textrm{\"{o}}dinger}(1952)}]{Schroed}
\bibinfo{author}{\bibfnamefont{E.}~\bibnamefont{Schr\textrm{\"{o}}dinger}},
  \bibinfo{journal}{Nature} \textbf{\bibinfo{volume}{169}},
  \bibinfo{pages}{538} (\bibinfo{year}{1952}).

\bibitem[{\citenamefont{Takabayasi}(1953)}]{Takabayasi}
\bibinfo{author}{\bibfnamefont{T.}~\bibnamefont{Takabayasi}},
  \bibinfo{journal}{Progr. Theor. Phys.} \textbf{\bibinfo{volume}{9}},
  \bibinfo{pages}{187} (\bibinfo{year}{1953}).

\bibitem[{\citenamefont{Akhmeteli}({\natexlab{b}})}]{Akhm18}
\bibinfo{author}{\bibfnamefont{A.}~\bibnamefont{Akhmeteli}},
  \emph{\bibinfo{title}{quant-ph/1006.2578}}.

\bibitem[{\citenamefont{Itzykson and Zuber}(1980)}]{Itzykson}
\bibinfo{author}{\bibfnamefont{C.}~\bibnamefont{Itzykson}} \bibnamefont{and}
  \bibinfo{author}{\bibfnamefont{J.-B.} \bibnamefont{Zuber}},
  \emph{\bibinfo{title}{{Quantum field theory}}}
  (\bibinfo{publisher}{McGraw-Hill}, \bibinfo{year}{1980}).

\bibitem[{\citenamefont{Akhmeteli}({\natexlab{c}})}]{AkhmeteliJMParx}
\bibinfo{author}{\bibfnamefont{A.}~\bibnamefont{Akhmeteli}},
  \emph{\bibinfo{title}{quant-ph/1008.4828}}.

\bibitem[{\citenamefont{Feynman and Gell-Mann}(1958)}]{Feygel}
\bibinfo{author}{\bibfnamefont{R.~P.} \bibnamefont{Feynman}} \bibnamefont{and}
  \bibinfo{author}{\bibfnamefont{M.}~\bibnamefont{Gell-Mann}},
  \bibinfo{journal}{Phys. Rev.} \textbf{\bibinfo{volume}{109}},
  \bibinfo{pages}{193} (\bibinfo{year}{1958}).

\bibitem[{\citenamefont{Gustavson et~al.}(2018)\citenamefont{Gustavson,
  Ovchinnikov, and Pogudin}}]{Gust}
\bibinfo{author}{\bibfnamefont{R.}~\bibnamefont{Gustavson}},
  \bibinfo{author}{\bibfnamefont{A.}~\bibnamefont{Ovchinnikov}},
  \bibnamefont{and} \bibinfo{author}{\bibfnamefont{G.}~\bibnamefont{Pogudin}},
  \bibinfo{journal}{Journal of Symbolic Computation}
  \textbf{\bibinfo{volume}{85}}, \bibinfo{pages}{128} (\bibinfo{year}{2018}).

\bibitem[{\citenamefont{Bogoliubov and Shirkov}(1980)}]{Bogo}
\bibinfo{author}{\bibfnamefont{N.~N.} \bibnamefont{Bogoliubov}}
  \bibnamefont{and} \bibinfo{author}{\bibfnamefont{D.~V.}
  \bibnamefont{Shirkov}}, \emph{\bibinfo{title}{{Introduction to the Theory of
  Quantized Fields\rm , 3rd ed.}}} (\bibinfo{publisher}{J. Wiley, New York},
  \bibinfo{year}{1980}).

\bibitem[{\citenamefont{Akhmeteli}({\natexlab{d}})}]{Akhm2015}
\bibinfo{author}{\bibfnamefont{A.}~\bibnamefont{Akhmeteli}},
  \emph{\bibinfo{title}{quant-ph/1502.02351}}.

\end{thebibliography}
\end{document}